\pdfoutput=1
\documentclass[groupedaddress,
preprint,
 amsmath,amssymb,
 aps,
 prd,
]{revtex4-2}

\usepackage{color}
\usepackage{graphicx}\usepackage{xspace}
\usepackage{multirow}
\usepackage{placeins}
\usepackage{tabularray}

\usepackage{hyperref}

\newcommand*{\MGMCatNLO}{\textsc{MadGraph5}\_aMC@NLO\xspace}
\newcommand*{\MadGraph}{\MGMCatNLO}

\newcommand*{\ssWWVBS}{\ensuremath{\mathrm{W}^\pm \mathrm{W}^\pm \mathrm{jj}}}
\newcommand*{\ssWWVBSLL}{\ensuremath{\mathrm{W_{\mathrm{L}}}^\pm \mathrm{W_{\mathrm{L}}}^\pm \mathrm{jj}}}
\newcommand*{\ssWWVBSLT}{\ensuremath{\mathrm{W_{\mathrm{L}}}^\pm \mathrm{W_{\mathrm{T}}}^\pm \mathrm{jj}}}
\newcommand*{\ssWWVBSTL}{\ensuremath{\mathrm{W_{\mathrm{T}}}^\pm \mathrm{W_{\mathrm{L}}}^\pm \mathrm{jj}}}
\newcommand*{\ssWWVBSTT}{\ensuremath{\mathrm{W_{\mathrm{T}}}^\pm \mathrm{W_{\mathrm{T}}}^\pm \mathrm{jj}}}
\newcommand*{\WZVBS}{\ensuremath{\mathrm{WZ jj}}}

\newcommand{\fastjet}{\textsc{FastJet}}

\begin{document}
\title{Sensitivity to longitudinal vector boson scattering and doubly-charged Higgs boson production in $\mathbf\ssWWVBS$ at future hadron colliders}

\author{Aram Apyan}
\author{Samuel Kelson}
\affiliation{Department of Physics, Brandeis University, Waltham MA 02453, USA}
\author{Chilufya Mwewa}
\author{Marc-Andr\'e Pleier}
\affiliation{Physics Department, Brookhaven National Laboratory, Upton, New York 11973-5000, USA}
\author{Luka Nedic}
\affiliation{Department of Physics, Oxford University, Oxford OX1 3PJ, UK}
\author{Karolos Potamianos}
\affiliation{Department of Physics, University of Warwick, Coventry; United Kingdom}
\date{\today}
\begin{abstract}
We study the sensitivity to longitudinal vector boson scattering at a 27, 50 and 100~TeV $pp$ collider using events containing two leptonically-decaying same-electric-charge $W$ bosons produced in association with two jets. The baseline FCC-hh detector parameterization within the Delphes framework is used under the assumption of fully efficient pile-up mitigation. A tightly constrained phase space with a dijet mass greater than 2 TeV is considered in order to suppress the impact of potential instrumental backgrounds. Based on this setup, the expected sensitivity to the production of longitudinally polarized same-sign $W$ boson pairs is evaluated. Additionally, expected limits are set on doubly charged Higgs bosons produced via vector boson fusion processes and decaying to same-sign $W$ boson pairs.
\end{abstract}

\maketitle
\pagebreak{}
\section{\label{sec:intro}Introduction}

Vector boson scattering (VBS) processes are important probes of the non-Abelian gauge structure of the electroweak (EW) interactions 
and the EW symmetry breaking mechanism. The unitarity of the tree-level amplitude of the scattering of longitudinally polarized gauge 
bosons at high energies~\cite{Veltman:1976rt,Lee:1977yc,Lee:1977eg} \emph{could} be restored by the Higgs-like boson observed at the 
CERN Large Hadron Collider (LHC) with a mass of about $125$ GeV~\cite{AtlasPaperCombination,CMSPaperCombination}. However, modifications
of the cross sections of processes involving scattering of longitudinally polarized gauge bosons with respect to the Standard Model (SM)
expectations are predicted in physics beyond the SM models via the presence of additional new resonances or via modifications of the 
Higgs boson couplings to gauge bosons~\cite{Espriu:2012ih,Chang:2013aya,Lee:2018fxj}.  

The VBS processes at proton-proton ($pp$) colliders are characterized by the presence of two gauge bosons in association with a forward/backward pair of jets. The goal of this paper is to study the prospects for measuring the polarized scattering of same-electric-charge (same-sign) \ssWWVBS~production at a future high energy $pp$ machine~\cite{FCC:2018vvp,CEPCStudyGroup:2018rmc}. In particular, our studies focus on future hadron colliders operating at a center-of-mass energy ($\sqrt{s}$) of 27, 50 and 100 TeV (corresponding respectively to the High-Energy LHC (HE-LHC) using advanced high-field dipole magnets, a Future Circular Collider (FCC)-like ring using LHC magnet technology, and to the envisioned FCC-hh collider~\cite{FCC:2018vvp}) with an integrated luminosity of 30 ab$^{-1}$. The leptonic decay mode of EW production of same-sign \ssWWVBS, where both W bosons decay into electrons or muons (including those coming from tau leptons) and their respective (anti-) neutrinos, is a promising final state as the ratio between the strong (QCD-induced) and electroweak production of \ssWWVBS~is small. The ATLAS and CMS Collaborations have measured EW \ssWWVBS~production at $\sqrt{s}=8$ and $13$ TeV~\cite{Aad:2014zda,Khachatryan:2014sta,Sirunyan:2017ret,Aaboud:2019nmv,Sirunyan:2020gyx, ATLAS:2023sua}. 

The W boson can be polarized either longitudinally ($\mathrm{W_{\mathrm{L}}}$) or transversely ($\mathrm{W_{\mathrm{T}}}$). This leads to 
three contributions, \ssWWVBSLL, \ssWWVBSLT~(which includes \ssWWVBSTL), and \ssWWVBSTT~to the overall \ssWWVBS~production. In this study, the candidate \ssWWVBS~events
contain exactly two leptons with the same electric charge, missing transverse momentum, and two jets with a large rapidity separation and a high dijet invariant
mass. The helicity eigenstates are defined in the WW center-of-mass reference frame which has a higher contribution from longitudinally polarized W bosons compared to the center-of-mass frame of the colliding partons. Studies of changes in polarization fractions and in kinematic distributions  arising from defining the helicity eigenstates in different reference frames at the LHC are reported in Ref.~\cite{Ballestrero:2020qgv}. A maximum-likelihood fit is performed using the distribution of events as a function of a Boosted Decision Tree (BDT) variable optimized to distinguish between signal and background processes that can mimic the signal signature, to extract the cross sections of the \ssWWVBSLL, \ssWWVBSLT, and \ssWWVBSTT~contributions.

The first measurements of the production cross sections of polarized scattering in the same-sign \ssWWVBS~process were performed at the LHC by the ATLAS and CMS Collaborations using data samples corresponding to integrated luminosities of about 140 fb$^{-1}$ at $\sqrt{s}=13$ TeV~\cite{ATLAS:2025wuw, CMS:2020etf}. An observed significance of 3.3 standard deviations for the production with at least one longitudinally polarized W boson was reported by the ATLAS Collaboration while an observed significance of 2.3 standard deviations was reported by the CMS Collaboration. There have been a number of studies focusing on techniques to maximize the sensitivity to the purely longitudinal contributions at the High-Luminosity LHC (HL-LHC)~\cite{Searcy:2015apa,Lee:2018xtt,Azzi:2019yne,Roloff:2021kdu}. The CMS Collaboration recently projected the result in Ref.~\cite{CMS:2020etf} to an integrated luminosity of $3000$ fb$^{-1}$,
expected at the end of the HL-LHC, to obtain an expected significance of about four standard deviations for 
\ssWWVBSLL~production~\cite{CMS-PAS-FTR-21-001}. Beyond the LHC, studies focusing on the prospects of measuring longitudinally polarized ZZ
scattering at a future high-energy muon collider were performed in Ref.~\cite{Yang:2021zak}.

Some models beyond the SM predict charged Higgs bosons that couple to $W$ and $Z$ bosons at tree level. One such model is the Georgi-Machacek (GM) model~\cite{GMModel}. This study sets expected limits on the production of doubly charged Higgs bosons produced via vector boson fusion processes and decaying to same-sign $W$ boson pairs in the context of the GM model as a benchmark for Beyond the SM (BSM) sensitivity. Both the CMS and ATLAS Collaborations have recently published limits on this process using data samples corresponding to 137 fb$^{-1}$~\cite{CMS-s10052-021-09472-3} and 139 fb$^{-1}$~\cite{ATLAS:2023sua}, respectively, at $\sqrt{s}=13$ TeV. The ATLAS Collaboration reported an excess of events corresponding to two and a half standard deviations for a doubly charged Higgs boson with a mass of 450 GeV.
 \section{\label{sec:mc}Event Simulation}
Signal events are modeled at leading order (LO) in QCD using the Monte Carlo event generator \MadGraph version 3.4.1 \cite{Madgraph:2014,BuarqueFranzosi:2019boy} with the next-to-next-to-LO (NNLO) hessian NNPDF3.1 parton distribution function (PDF) set \cite{PDFSets:2013} interfaced with PYTHIA version 8.306 \cite{PYTHIA8:2015} for parton showering. Events where either both W bosons are
longitudinally polarized (\ssWWVBSLL~events), both W bosons are transversely polarized (\ssWWVBSTT~events), or those where one of the W bosons is longitudinally
polarized and the other is transversely polarized (\ssWWVBSLT~events) are generated separately. 

Recent calculations of the fixed-order next-to-LO (NLO) EW and QCD corrections for the production of polarized EW \ssWWVBS~at $\sqrt{s}=13$ TeV show a significant reduction of the LO cross sections~\cite{Denner:2024tlu}. The corrections reduce the cross sections by approximately 10\% (15\%) for the \ssWWVBSLL~(\ssWWVBSTT) process in a typical fiducial region used by the ATLAS and CMS Collaborations. The corrections are larger in high-energy tails of distributions, but are not available at center-of-mass energies considered here and the investigation of the impact of the higher order corrections for these processes is beyond the scope of this work. However, the EW corrections for the unpolarized \ssWWVBS~production have been computed at $\sqrt{s}=27$ TeV~\cite{Azzi:2019yne} and are only a few percent larger compared to predictions at 13 TeV. 

Background processes considered in this analysis include the production of \ssWWVBS~ via the strong interaction, the production of
\WZVBS~via the electroweak and strong interactions, and the production of $\mathrm{t}\ell^+\ell^- \mathrm{j}$, referred to as tZq. These processes are simulated with \MadGraph~as well. 

The FCC-hh detector requires unprecedented timing and spatial resolution to mitigate the effects of up to 1000 $pp$ interactions per bunch crossing (pile-up). High-precision tracking and calorimetry systems in the forward region will extend the pseudorapidity coverage up to $|\eta| < 6$~\cite{FCC:2018vvp, Aleksa:2019pvl}. The performance of the baseline FCC-hh detector is parameterized within the Delphes framework~\cite{deFavereau:2013fsa, DELPHES:FCChh-card}, using dedicated simulations to determine the electromagnetic and hadronic calorimeter resolutions, as well as the tracking performance~\cite{Selvaggi:2717698}. The reconstruction and identification efficiencies, along with the energy and momentum resolutions for electrons and muons, closely follow the performance achieved at the LHC. A simplified particle-flow algorithm is employed to form particle candidates, which are then used for jet clustering and the calculation of the missing transverse momentum. This FCC-hh benchmark Delphes detector card~\cite{DELPHES:FCChh-card} is used to simulate detector effects for all signal and background samples considered in this study. A detailed description of the detector parameters can be found in Ref.~\cite{Selvaggi:2717698}.

It should be noted that pile-up effects are not included in the Delphes FCC-hh detector description and are not part of the event simulation. For the purposes of this study, it is assumed that pile-up can be mitigated in future analyses by using precise timing detectors and advanced pile-up rejection algorithms.

Jets are clustered from the reconstructed objects using \fastjet~\cite{Cacciari:2011ma} with the anti-kt clustering algorithm~\cite{Cacciari:2008gp}, and a distance parameter of 0.4.
For the validation of the signal samples, an inclusive \ssWWVBS~ electroweak sample was generated using the same generator and settings as well as Delphes for detector effects, and good agreement was observed with respect to the sum of the polarized signal samples.  

Detector-specific backgrounds such as those in which lepton charge is misidentified or those in which jets mimic leptons were not considered in this analysis as they are highly detector-specific.

 \section{\label{sec:sel}Event Selection}

Signal (\ssWWVBS) events are characterized by the presence of two high transverse momentum ($p_T$) same-charge leptons 
(electrons or muons, denoted by $\ell$), large missing transverse momentum ($E^{miss}_T$)
and two forward/backward jets with large di-jet invariant mass ($M_{jj}$). In this analysis, events with $\tau$ leptons are only considered when the $\tau$ decays leptonically (to $e$ or $\mu$). A combination of single and double lepton triggers are used in ATLAS and CMS EW same-sign WW measurements~\cite{CMS:2020etf} to ensure a trigger efficiency above
99\% for events in the analysis fiducial region. For the results reported in this study, it is assumed that a similar trigger selection efficiency can be achieved with the future FCC detectors.

For simplicity, the event selection was optimised for $\sqrt{s} = 100$~TeV to maximize the background rejection and enhance the signal and then applied to all $\sqrt{s}$ values under study. The di-lepton invariant mass ($M_{\ell\ell}$) was required to be at least 60 GeV in order to minimize the background contribution from low mass Drell-Yan processes. To eliminate leptons from Z bosons, the invariant mass of a pair of electrons or muons was required to be away from the Z mass by at least 10 GeV. 

A large di-jet invariant mass of at least 2 TeV was required in order to satisfy the VBS topology. This requirement was significantly more restrictive compared to the typical $M_{jj}>500$ GeV requirement used in the corresponding LHC measurements~\cite{Sirunyan:2020gyx, ATLAS:2023sua, ATLAS:2025wuw, CMS:2020etf}. While the more restrictive requirement used in this paper is not optimal for the sensitivity to \ssWWVBSLL~production, it greatly reduces the contributions from the  backgrounds such as those in which lepton charge is misidentified or those in which jets mimic leptons. These highly detector-specific background processes are typically estimated using data-driven methods and have negligible contributions for events with $M_{jj}>2$ TeV~\cite{Sirunyan:2020gyx, ATLAS:2023sua} at the LHC. For the purposes of this study, it is assumed that the contribution of these background processes, primarily arising from events where a hadron or a lepton from a hadron decay passes the lepton selection~\cite{ATLAS:2022swp}, remains negligible at higher center-of-mass energies considered here.

Table~\ref{tab:eventselectionTable} gives a summary of all object and event selection cuts applied in this analysis. 
The distributions of events as a function of the separation in azimuthal angle $\phi$ between the two leading jets in $p_T$ ($\Delta \phi_{jj}$) are shown in Figure~\ref{fig:preFitDPhijj} for all the three center-of-mass energies.

\begin{table}[h!]
\centering
\begin{tabular}{l|l}
\hline \hline
Selection type & Requirement \\
\hline \hline
Number of leptons & Exactly 2  same-charge leptons\\
Lepton $p_T$ & $p_T \geq 15$ GeV \\
Number of jets & $\geq 2$ \\
Jet $p_T$ & $p_T \geq 50$ GeV \\
Di-lepton invariant mass & $M_{ll} \geq 60$ GeV \\
$Z-$veto & $|M_{ll} - M_Z| > 10$ GeV \\
Di-jet invariant mass & $M_{jj} \geq 2$ TeV \\
Missing transverse momentum & $E^{miss}_T \geq 50$ GeV \\
\hline \hline
\end{tabular}
\caption{Selection criteria used for $W^\pm W^\pm jj$ events.}
\label{tab:eventselectionTable}
\end{table}

 \section{\label{sec:theory}Theory uncertainties}
There are three primary sources of theoretical uncertainties associated with perturbative QCD matrix-element calculations. The first and largest is due to the missing higher-order terms in perturbative expansions. To estimate this uncertainty, samples are generated with variations of the renormalization ($\mu_R$) and factorization $(\mu_F$) scales. Pairwise variations of up to a factor of 2 are performed in the up and down direction $\{\mu_R,\mu_F\} \times \{0.5,0.5\}, \{1,0.5\}, \{0.5,1\}, \{1,1\}, \{2,1\}, \{1,2\}, \{2,2\}$, with the final uncertainty evaluated by taking the envelope of all seven-point scale variations.
In addition to the scale variations, PDF and $\alpha_S$ uncertainties also contribute to the theoretical uncertainty. The theoretical uncertainty associated with the PDF is evaluated by varying the internal parameters following the prescription from \cite{Butterworth_2016}. Finally, the third source of theoretical uncertainty is associated with the determination of the strong coupling constant, $\alpha_S$. The coupling constant is determined experimentally using fixed-order calculations in perturbation theory. 
The $\alpha_S$ uncertainties are evaluated at two different $\alpha_S$ values and the uncertainty is calculated following the recommendations of \cite{Butterworth_2016}. 

\begin{table}[htbp]
\begin{center}
\begin{tabular}{|c|c|c|c|c|c|c|c|}
\hline 
  $\sqrt{s}$ (TeV)   & $W_{L}W_{L}$      & $W_{L}W_{T}$      & $W_{T}W_{T}$      & WW QCD      & $\;$ tZq $\;$     & WZ EW      & WZ QCD \\ 
\hline 
 27 & 14.7 & 15.4 & 15.8 & 23.7 & 10.4 & 16.3 & 52.0 \\ 
 50 & 10.8 & 9.99 & 10.1 & 18.4 & 10.3 & 11.1 & 37.6 \\ 
 100 & 4.92 & 5.32 & 5.40 & 13.0 & 10.4 & 5.48 & 30.8 \\
\hline 
\end{tabular} 
\caption{Relative percentage effect on the event yields of the theory systematics for each sample.\label{theorysystematics}} 
\end{center}
\end{table} 

Table \ref{theorysystematics} summarizes the effect of theory uncertainties on their respective sample as a relative percentage of the total event yield. The systematics have the largest impact on the $27$ TeV samples, with the effects decreasing with center-of-mass energy. The largest source of uncertainty for all center-of-mass energies arises from the scale uncertainties on the QCD-induced WZ background.

 \section{\label{sec:analysis}Sensitivity to $W^\pm W^\pm jj$ polarizations}

To separate the three \ssWWVBS~polarizations from background processes, two Boosted Decision Trees (BDTs) were trained using XGBoost~\cite{XGBoost}. The first BDT separates the \ssWWVBS~signal from the background processes (referred to as SB BDT) and the second BDT separates the three \ssWWVBS~polarization states (referred to as Pol BDT).
Several kinematic variables were used as input to the BDTs, as can be seen in Table~\ref{tab:BDTVariables}. This subset of variables for training was chosen from a larger set based on performance. Additionally, a correlation coefficient matrix was generated. The top $15$ of the initial $25$ variables were kept while avoiding keeping multiple variables with high correlations with respect to each other.
Hyperparameters, including the learning rate, maximum depth, subsample ratio, and number of estimators, were tuned via a random grid search algorithm~\cite{scikit-learn}. Simulated MC events were split into training and testing samples used for validation.

\begin{table}[h!]
\centering
\resizebox{0.86\textwidth}{!}{\begin{tblr}{Q[2.5cm,m]|Q[10cm,valign=m]|Q[1cm,valign=m,c]|Q[1cm,valign=m,c]}
\hline \hline
Variable & Description & SB & Pol\\
\hline \hline
$|\Delta \eta _{\ell\ell}|$ & Difference in rapidity of the leading and sub-leading leptons &  & \checkmark \\
$|\Delta \phi _{\ell_0E^T_{\text{miss}}}|$ & Difference in azimuthal angles of the leading lepton and missing transverse energy &  & \checkmark \\
$|\Delta \phi _{\ell_1 E^T_{\text{miss}}}|$ & Difference in azimuthal angles of the sub-leading lepton and missing transverse energy & \checkmark & \checkmark \\
$|\Delta \phi _{\ell_0 \ell_1}|$ & Difference in azimuthal angles of the leading and sub-leading leptons &  & \checkmark \\
$|\Delta \phi _{\ell\ell E^T_{\text{miss}}}|$ & Difference in azimuthal angles of the dilepton system and missing transverse energy & \checkmark &  \\
$|\Delta R _{j_1\ell_1}|$ & Distance between the leading jet and leading lepton &  & \checkmark \\
$|\Delta R _{j_2\ell_2}|$ & Distance between the sub-leading jet and sub-leading lepton & \checkmark &  \\
$|\Delta R _{jj}|$ & Distance between the leading jet and sub-leading jet & \checkmark & \checkmark \\
$E^T_{\text{miss}}$ & Missing transverse energy & \checkmark & \checkmark \\
$m _{jj}$ & Mass of the dijet system & \checkmark &  \\
$m _{\ell\ell}$ & Mass of the dilepton system & \checkmark & \checkmark \\
$p _{T j_1}$ & Transverse momentum of the leading jet & \checkmark &  \\
$p _{T j_2}$ & Transverse momentum of the sub-leading jet & \checkmark &  \\
$p _{T j_3}$ & Transverse momentum of the 3rd leading jet & \checkmark & \checkmark \\
$p _{T jj}$ & Transverse momentum of the dijet system &  & \checkmark \\
$p _{T \ell_1}$ & Transverse momentum of the leading lepton & \checkmark & \checkmark \\
$p _{T \ell_2}$ & Transverse momentum of the sub-leading lepton &  & \checkmark \\
$p _{T \text{rel}}$ & Transverse momentum ratios for leptons and jets & \checkmark & \checkmark \\
$\sum \eta _\ell$ & Sum of rapidity of all leptons & \checkmark &  \\
$\text{Zeppenfeld}_{\ell1}$ & Zeppenfeld variable \cite{Zeppenfeld:1996} for the leading lepton & \checkmark & \checkmark \\
$\text{Zeppenfeld}_{\ell2}$ & Zeppenfeld variable for the sub-leading lepton & \checkmark &  \\
\hline \hline
\end{tblr}}
\caption{Variables, their descriptions, and which BDT they were used in.}
\label{tab:BDTVariables}
\end{table}

The Pol BDT is a multi-class classifier consisting of three outputs. Each output corresponds to a classification probability associated with one of the polarization states. The SB BDT is a binary classifier, with a single output giving the classification probability of an inclusive \ssWWVBS~event. The two BDTs are combined within the analysis framework to generate a three-dimensional histogram. The $x$ and $y$ dimensions present the predicted probability of observing a longitudinal-longitudinal and transverse-transverse event coming from the output of the polarization BDT, and the $z$ dimension presents the probability of an inclusive \ssWWVBS~signal event coming from the SB BDT. As the sum of the predicted probability of the Pol BDT is equal to one, only two outputs are necessary for complete sensitivity to polarization. The three-dimensional BDT is ``unrolled" to form a one-dimensional histogram. The unrolling translates each individual bin in the three-dimensional histogram to an individual bin in a one-dimensional histogram.

\begin{figure}[h!]
\centering
\includegraphics[width=.32\linewidth]{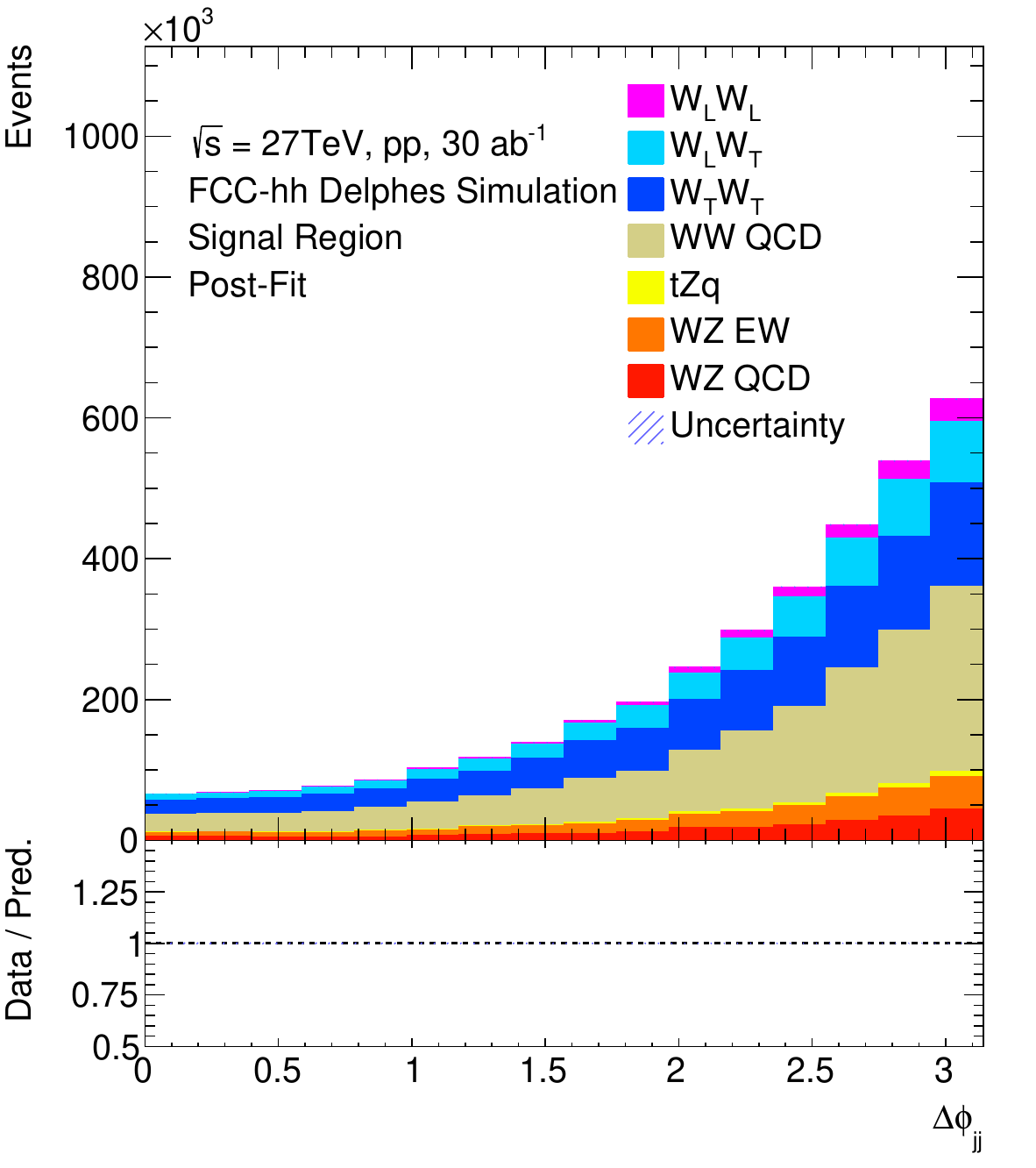}
\includegraphics[width=.32\linewidth]{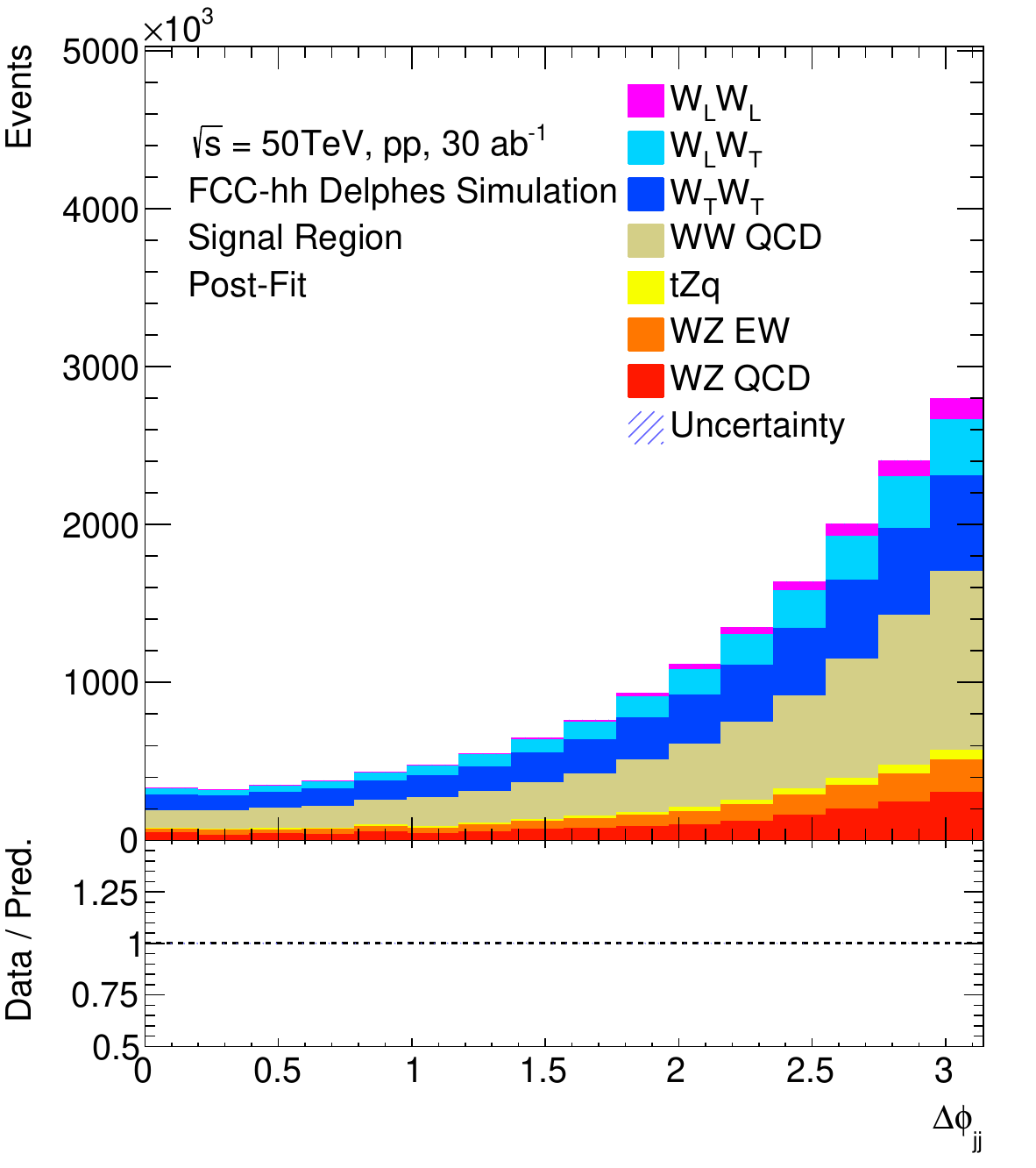}
\includegraphics[width=.32\linewidth]{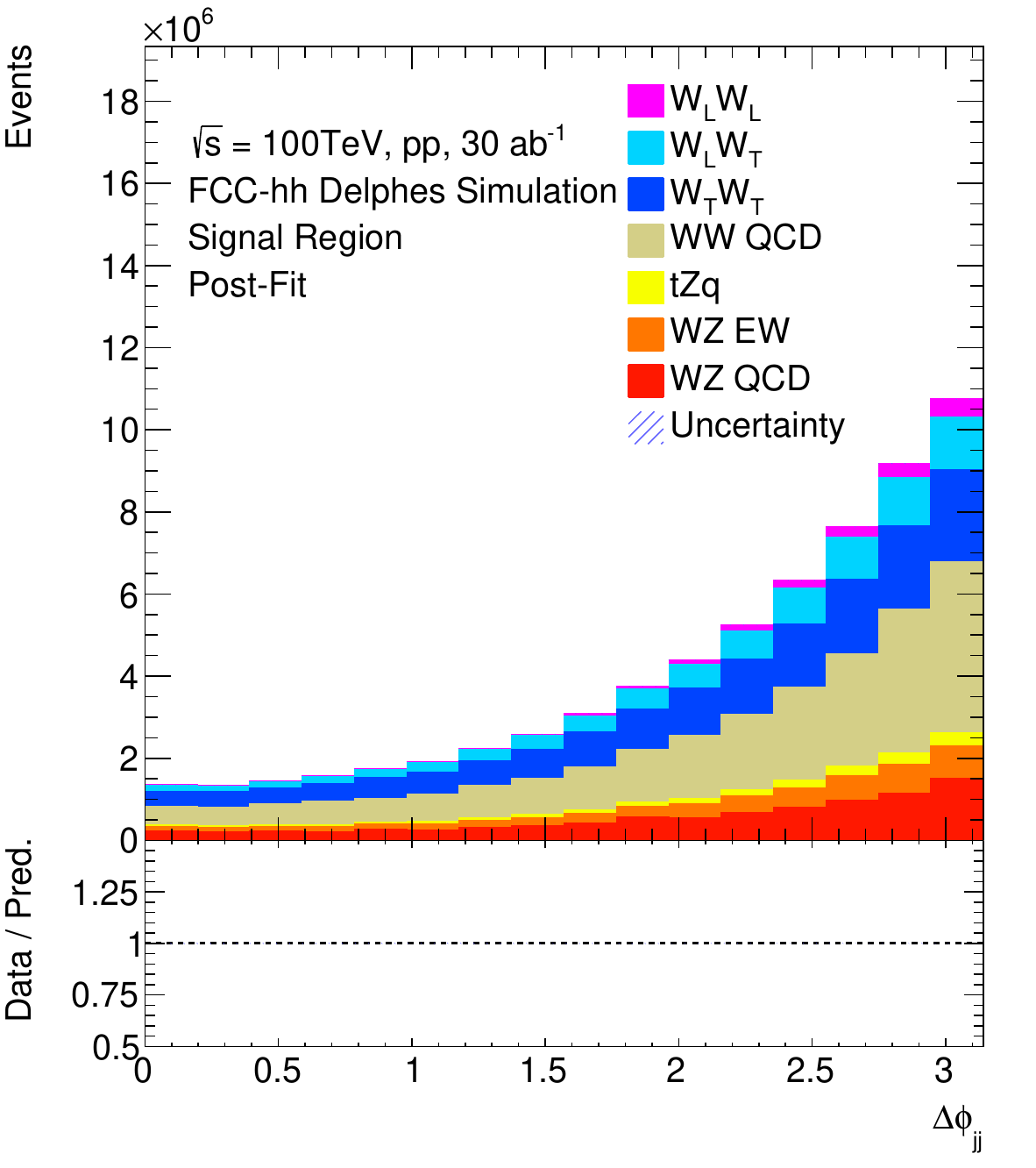}
\caption{The post-fit distribution of events as a function of $\Delta \phi_{jj}$.}
\label{fig:preFitDPhijj}
\end{figure}

\begin{figure}[h!]
\centering
\includegraphics[width=.32\linewidth]{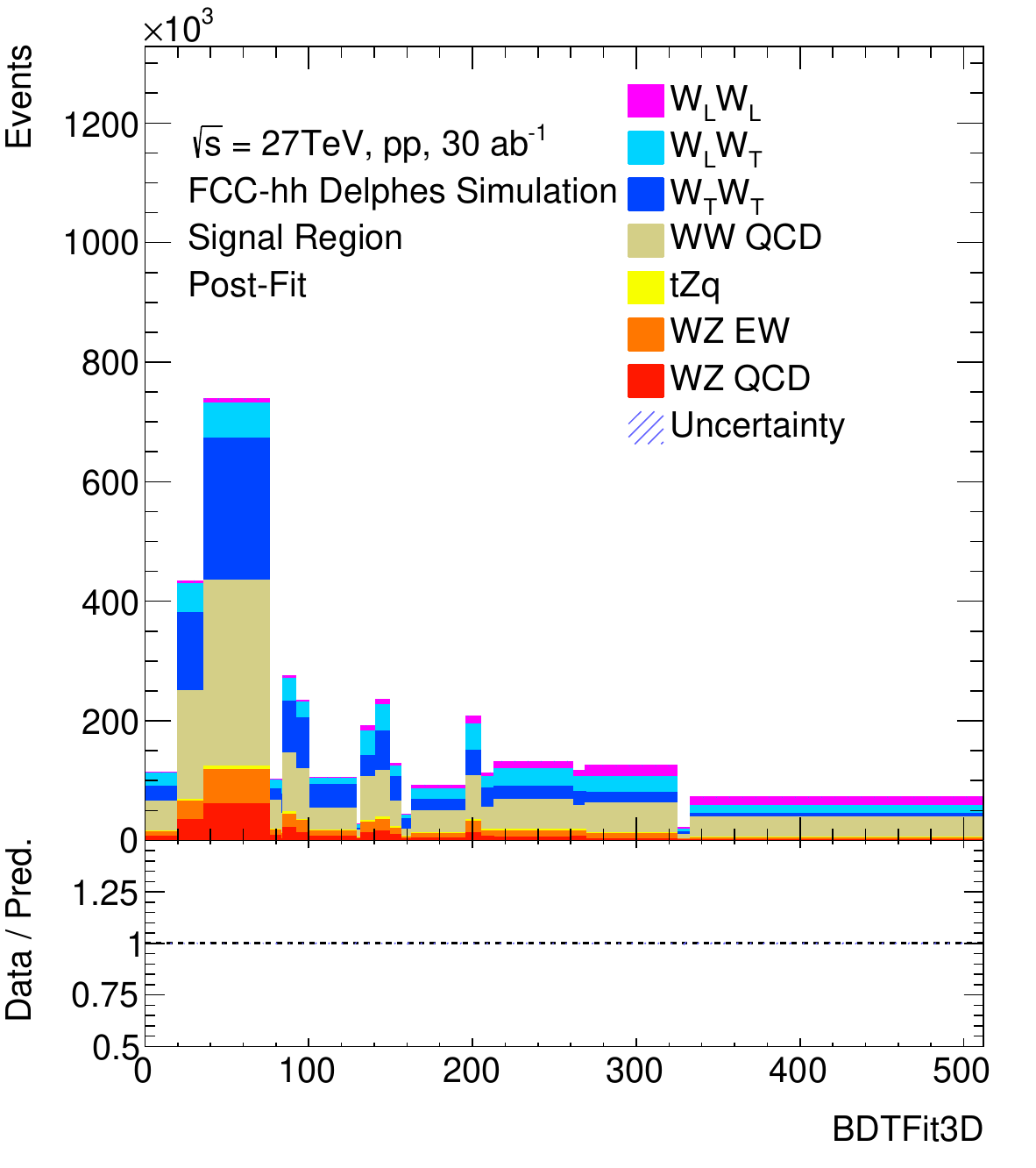}
\includegraphics[width=.32\linewidth]{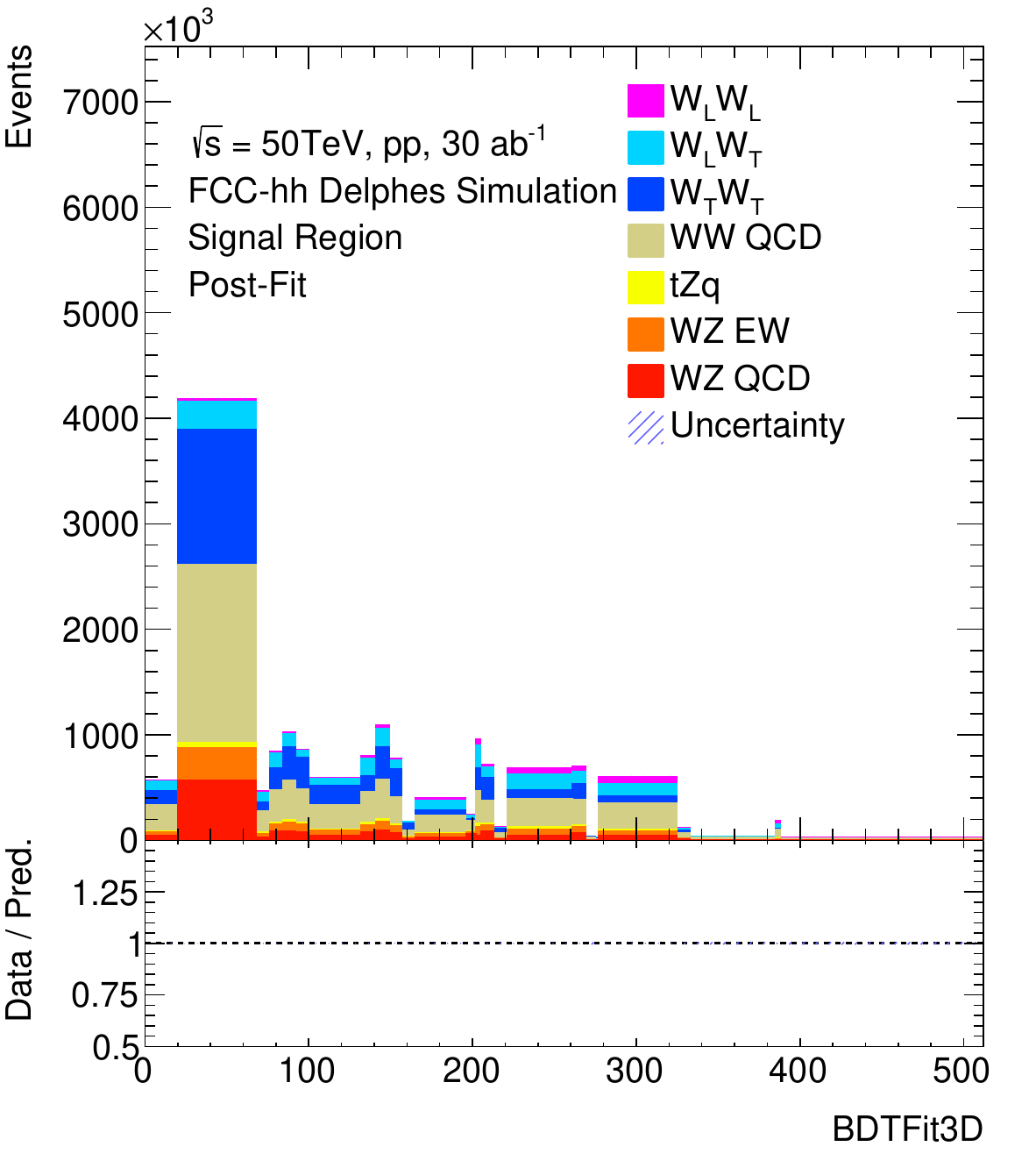}
\includegraphics[width=.32\linewidth]{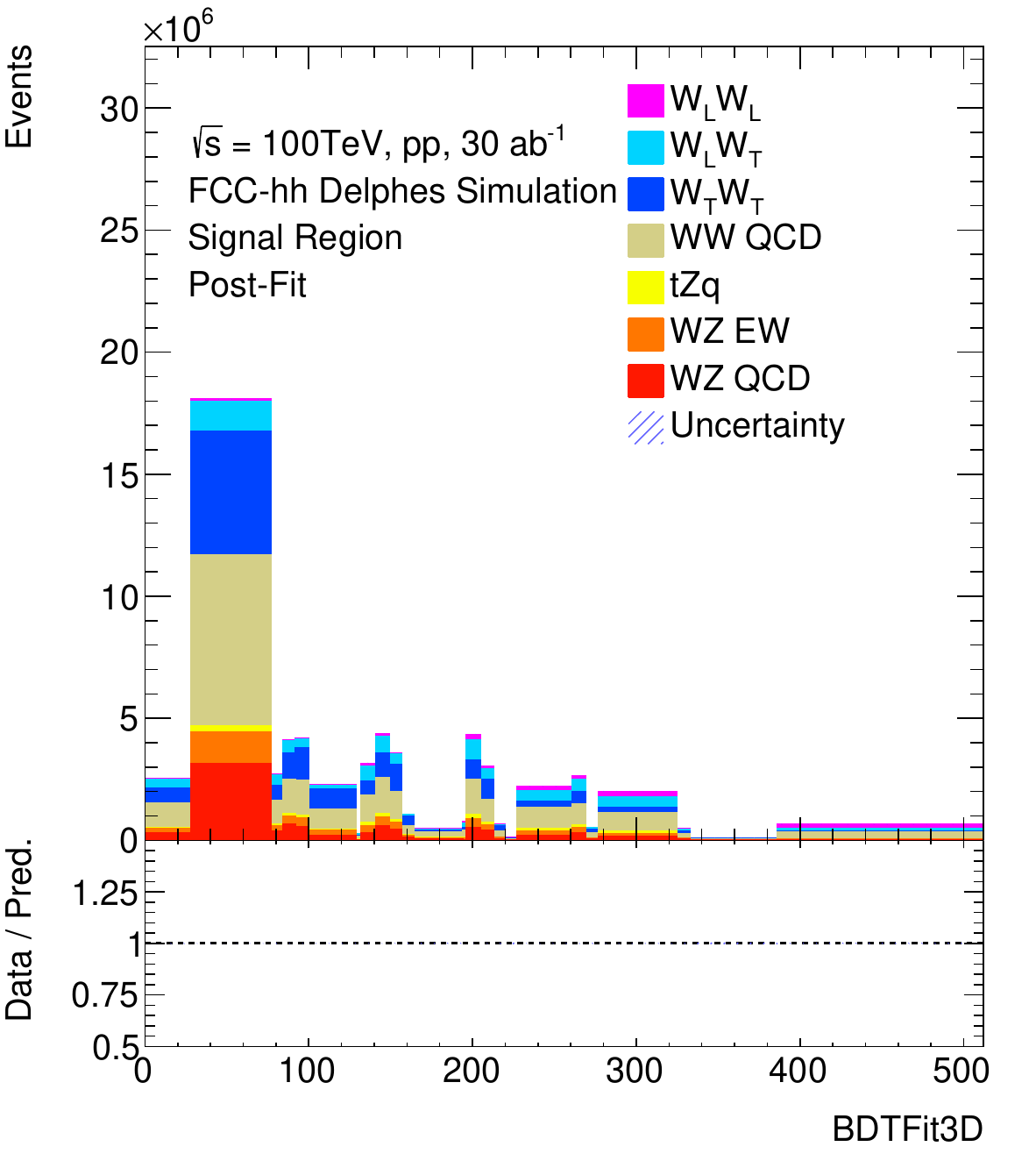}

\caption{ The post-fit distribution of events as a function of the unrolled BDT (BDTFit3D).}
\label{fig:preFitBDT}
\end{figure}

The expected sensitivity to the \ssWWVBS~polarization contributions is estimated using a profile likelihood fit as implemented in the
RooFit~\cite{Verkerke:2003ir} and RooStats~\cite{Moneta:2010pm} packages. We consider three signal strength parameters, $\mu_{LL}$, 
$\mu_{LT}$ and $\mu_{TT}$ corresponding to each of the \ssWWVBS~polarization states. In addition to the theoretical uncertainties, we assume a normalization uncertainty of $2\%$ for all the processes due to uncertainties in the integrated luminosity. The uncertainties associated with the limited number of simulated events are also included in the profile likelihood fit. Other sources of detector-specific experimental systematic uncertainties are not considered in this study. In the corresponding LHC measurements, the leading experimental systematic uncertainties arise from the limited number of data events in control regions used to constrain background processes~\cite{ATLAS:2025wuw}. It is assumed that these uncertainties can be significantly reduced due to the larger available data sample at the FCC-hh.

The sensitivity is measured for two distributions: $\Delta \phi_{jj}$ and the three-dimensional BDT. The $\Delta \phi_{jj}$ distribution is used as a baseline benchmark for the sensitivity to polarization and as a comparison to illustrate the improvements when applying multivariate analysis techniques such as BDTs. Before fitting, the BDT distributions are also subject to a binning optimization procedure. The procedure requires a fixed minimum of background and signal events per bin, as well as a minimum of 200 events per background. The procedure then iteratively tests merged bin configurations and compares each configuration by calculating the significance before and after merging. The binning configuration with the smallest number of bins, with the best significance, is taken as final. The significance is estimated as described in \cite{ATL-PHYS-PUB-2020-025}. The post-fit distributions for $\Delta \phi_{jj}$ and the BDT following the binning optimization procedure are shown in Figures~\ref{fig:preFitDPhijj} and \ref{fig:preFitBDT}.

\begin{table}[h!]
\begin{tabular}{|c|c|c|c|}
\hline\hline
\multirow{2}{*}{Polarization} & \multicolumn{3}{|c|}{Signal Strength: BDT} \\
\cline{2-4}
  & $\sqrt{s} = 27$~TeV & $\sqrt{s} = 50$~TeV & $\sqrt{s} = 100$~TeV \\ \hline \hline
$\mu_{LL}$ & $1 \pm 0.22$ & $1 \pm 0.21$ & $1 \pm 0.15$\\ \hline
$\mu_{LT}$ & $1 \pm 0.13$ & $1 \pm 0.095$ & $1 \pm 0.096$\\ \hline
$\mu_{TT}$ & $1 \pm 0.13$ & $1 \pm 0.085$ & $1 \pm 0.045$\\ \hline

 & \multicolumn{3}{|c|}{Signal Strength: $\Delta\phi_{jj}$} \\
 \hline
 $\mu_{LL}$ & $1 \pm 1.02$ & $1 \pm 0.62$ & $1 \pm 0.40$\\ \hline
$\mu_{LT}$ & $1  \pm 0.45$ & $1 \pm 0.42$ & $1 \pm 0.14$\\ \hline
$\mu_{TT}$ & $1 \pm 0.33$ & $1 \pm 0.26$ & $1 \pm 0.12$\\ \hline
\hline
\end{tabular}
\caption{\label{tab:normFactors} Signal strengths $\mu$ after the $\Delta\phi_{jj}$ and BDT fit for each \ssWWVBS~polarization state and all $\sqrt{s}$ values under study. }
\end{table}

Table~\ref{tab:normFactors} shows the signal strengths $\mu$, defined as the ratio of the fitted to the expected SM cross-section, and the
corresponding uncertainties for each \ssWWVBS~polarization state for all the considered $\sqrt{s}$ values. The values are expected to be unity because the
Asimov dataset~\cite{Cowan:2010js} is used in the fit. The breakdown of the uncertainties on the purely longitudinal scattering signal strength parameter is shown in Equation~\ref{eqn:breakdown} where ``stat" refers to the contribution of the statistical uncertainty to the overall uncertainty and ``syst" refers to the contribution of the systematic uncertainties, which include theoretical and luminosity uncertainties, as well as statistical uncertainties arising from limited number of MC events.

\begin{align}
\begin{split}
    \mu_{LL}^{27 \text{ TeV}} &= 1.00 \pm 0.22 = 1 \pm 0.17 \text{ (stat.)} \pm 0.14 \text{ (syst.)} \\
    \mu_{LL}^{50 \text{ TeV}} &= 1.00 \pm 0.21 = 1 \pm 0.18 \text{ (stat.)} \pm 0.11 \text{ (syst.)} \\
    \mu_{LL}^{100 \text{ TeV}} &= 1.00 \pm 0.15 = 1 \pm 0.12 \text{ (stat.)} \pm 0.09 \text{ (syst.)}
\end{split}
\label{eqn:breakdown}
\end{align}

The first detailed studies of \ssWWVBSLL~production at a future high energy $pp$ collider, with an integrated luminosity of 30 ab$^{-1}$, indicate that cross section measurements should be possible with a relative precision of around 22\% at $\sqrt{s} = 27$~TeV,  21\% at $\sqrt{s} = 50$~TeV, and 15\% at $\sqrt{s} = 100$~TeV.  A conservative event selection, with a dijet mass greater than 2 TeV, is considered in order to suppress the impact of potential instrumental backgrounds and pileup. Projections from the CMS Collaboration report an expected relative precision of about 25\% by the end of the HL-LHC program~\cite{CMS-PAS-FTR-21-001}.

\FloatBarrier

\section{Limits on doubly-charged Higgs boson production}

Charged Higgs bosons with couplings to W and Z bosons are included at tree level in extended Higgs sectors with additional isotriplet scalar fields~\cite{LHCHXSWG-2015-001}. In this study, we use the same-sign WW signal region described in Section \ref{sec:sel} to set limits on doubly charged Higgs bosons produced via Vector Boson Fusion (VBF). This is done in the context of the Georgi-Machacek model~\cite{GMModel} in which two isospin triplet scalar fields are added to the SM Higgs doublet. One of the constituents of the scalar potential is a quintuplet of Higgs bosons; $H^{\pm\pm}_5, H^{\pm}_5$, and $H^0_5$ of common mass, $m_{H^{\pm\pm}_5}$, that couple to $W$ and $Z$ bosons. Our interpretation uses the H5 plane benchmark hypothesis of the GM model which was developed by the Higgs cross section working group at the Large Hadron Collider (LHC)~\cite{LHCHXSWG-2015-001}. In this benchmark, doubly charged Higgs bosons ($H^{\pm\pm}_5$) decay to pairs of same-sign W bosons with a branching fraction of $100\%$. 
The production and decay of the $H^{\pm\pm}_5$ states depend on two parameters; the mass, $m_{H^{\pm\pm}_5}$, and the $\sin\theta_H$ (or $s_H$) parameter characterizing the contribution of isospin triplet scalar fields to the W and Z boson masses. The decay widths are proportional to $s_H^2$.

Signal samples were simulated using \MadGraph version 3.4.1 \cite{Madgraph:2014,BuarqueFranzosi:2019boy} at LO with the NNLO hessian NNPDF3.1 PDF set \cite{PDFSets:2013} interfaced with PYTHIA version 8.306 \cite{PYTHIA8:2015} for parton showering. The Delphes~\cite{deFavereau:2013fsa} program, using a generic FCC detector card~\cite{DELPHES:FCChh-card}, was then used to simulate detector effects. Five $H^{\pm\pm}_5$ mass points were simulated: 800 GeV, 900 GeV, 1000 GeV, 2000 GeV, and 3000 GeV. The high-mass region was selected due to the stringent limits already established by LHC experiments in the low mass region. Since $H^{\pm\pm}_5$ decay widths are proportional to $m_{H^{\pm\pm}_5}$ and $s_H^2$, $s_H$ values were set to 0.5 for the 800 GeV mass point, and to 0.25 for the higher masses.

To extract the limits, we performed a binned maximum-likelihood fit to the transverse mass ($m_T$) distribution of the di-lepton and $E^T_{\text{miss}}$ system which effectively distinguishes between the resonant signal and background processes. In addition to statistical uncertainties, systematic uncertainties, which include a 2$\%$ uncertainty on the luminosity, MC statistical uncertainties arising from a limited number of MC events, as well as theory uncertainties described in Section \ref{sec:theory}, were considered in this fit. Limits were extracted at $95\%$ CL using the $CL_s$ technique \cite{CLsTechnique}.

\begin{figure}[h!]
\centering
\includegraphics[width=.32\linewidth]{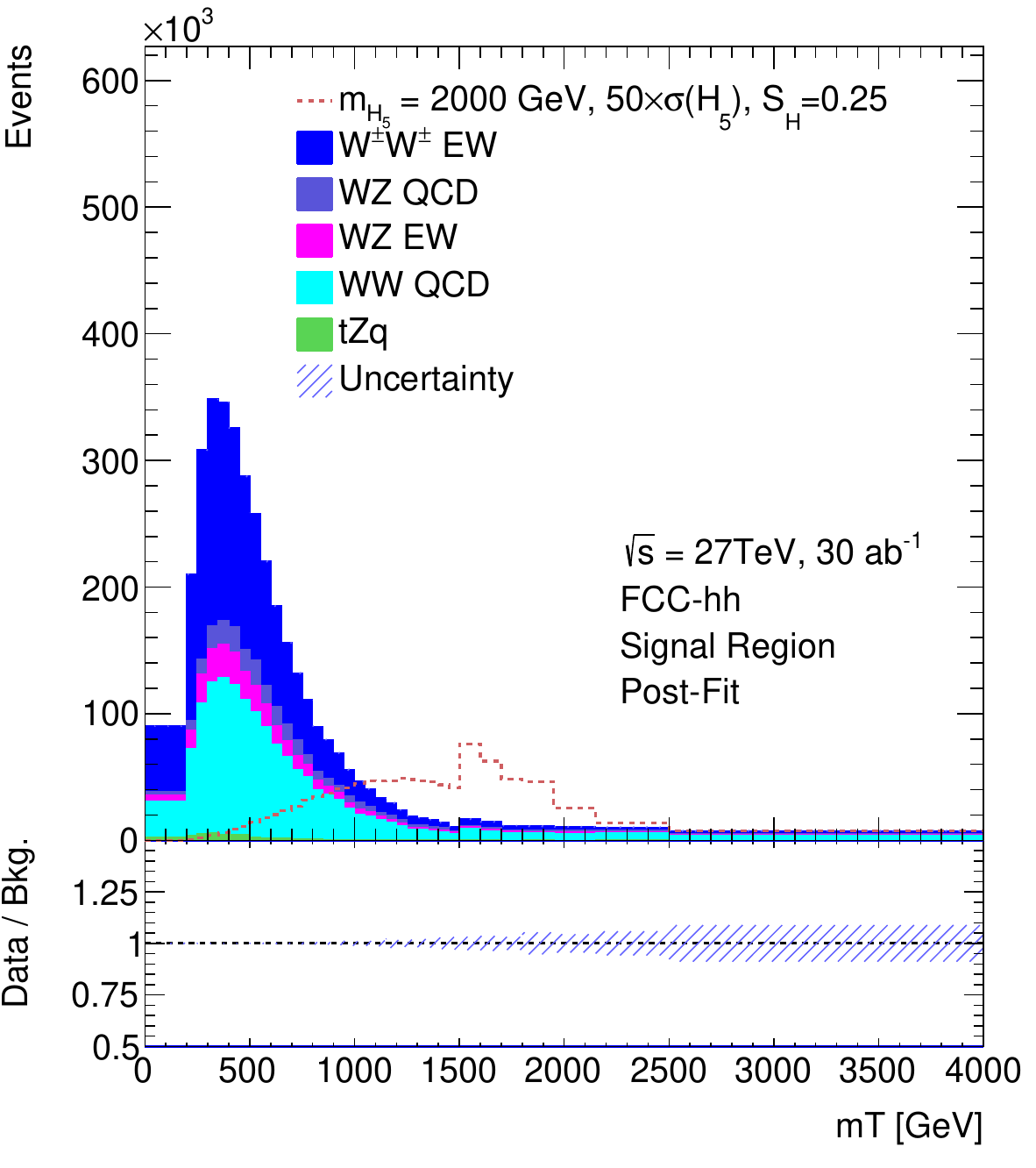}
\includegraphics[width=.32\linewidth]{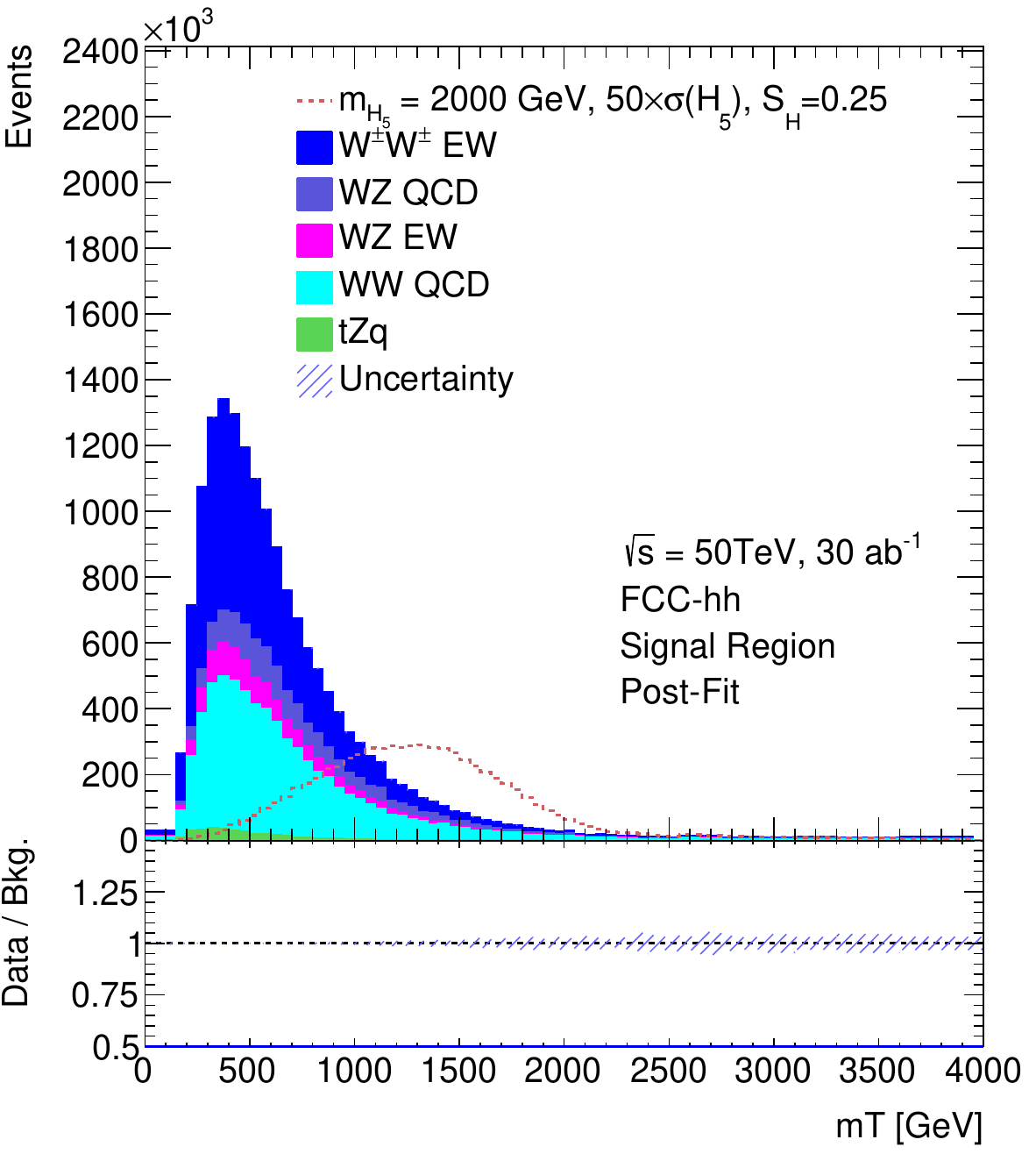}
\includegraphics[width=.32\linewidth]{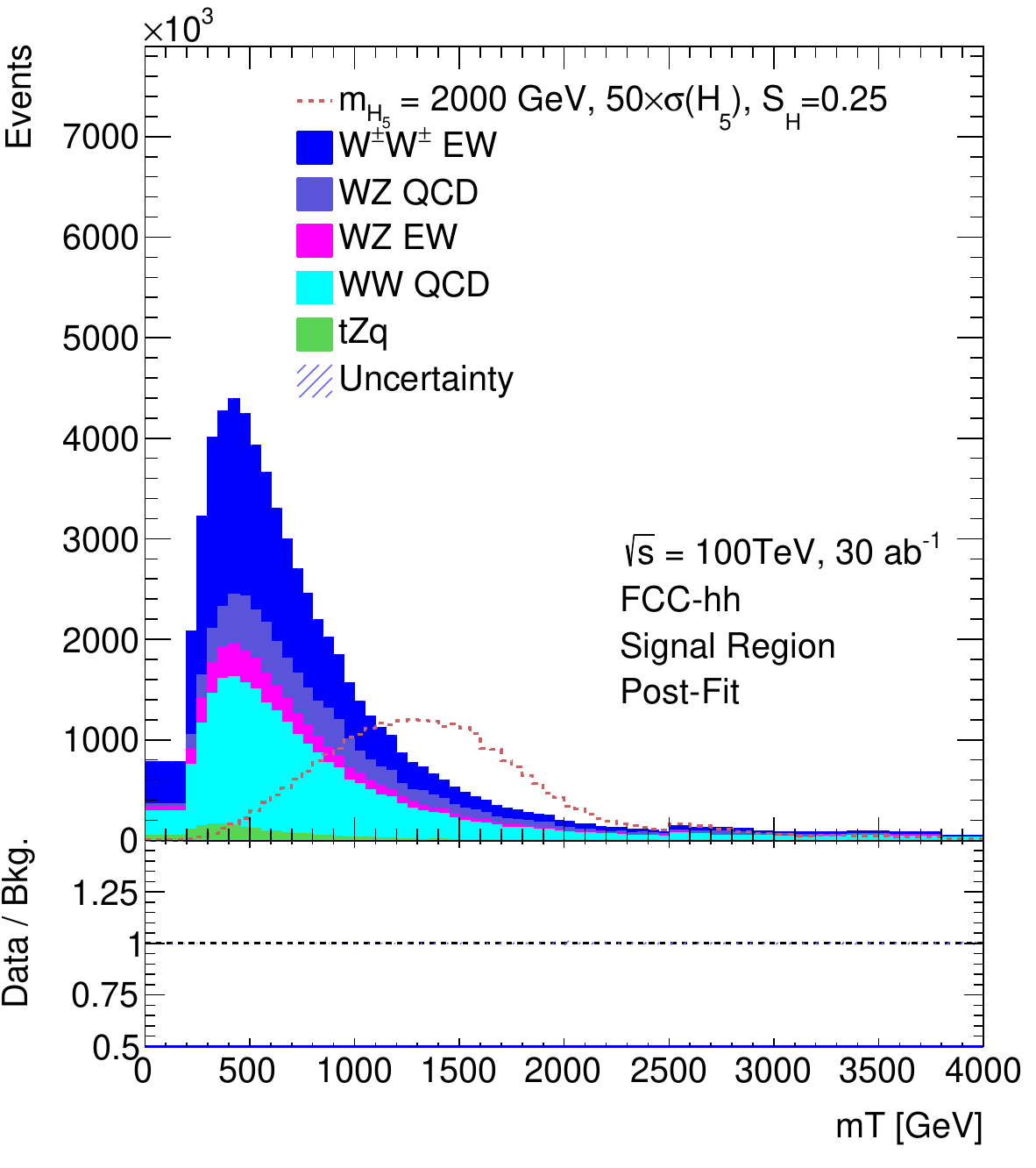}
\caption{Post-fit $m_T$ distributions for the 2000 GeV $H^{\pm\pm}_5$ signal at the three FCC-hh center-of-mass energy scenarios; 27 TeV (left), 50 TeV (center) and 100 TeV (right). The bins are optimized to ensure a minimum of 100 background events per bin, resulting in variable bin sizes.}
\label{Hpp:postfit}
\end{figure}

\begin{figure}[h!]
\centering
\includegraphics[width=.49\linewidth]{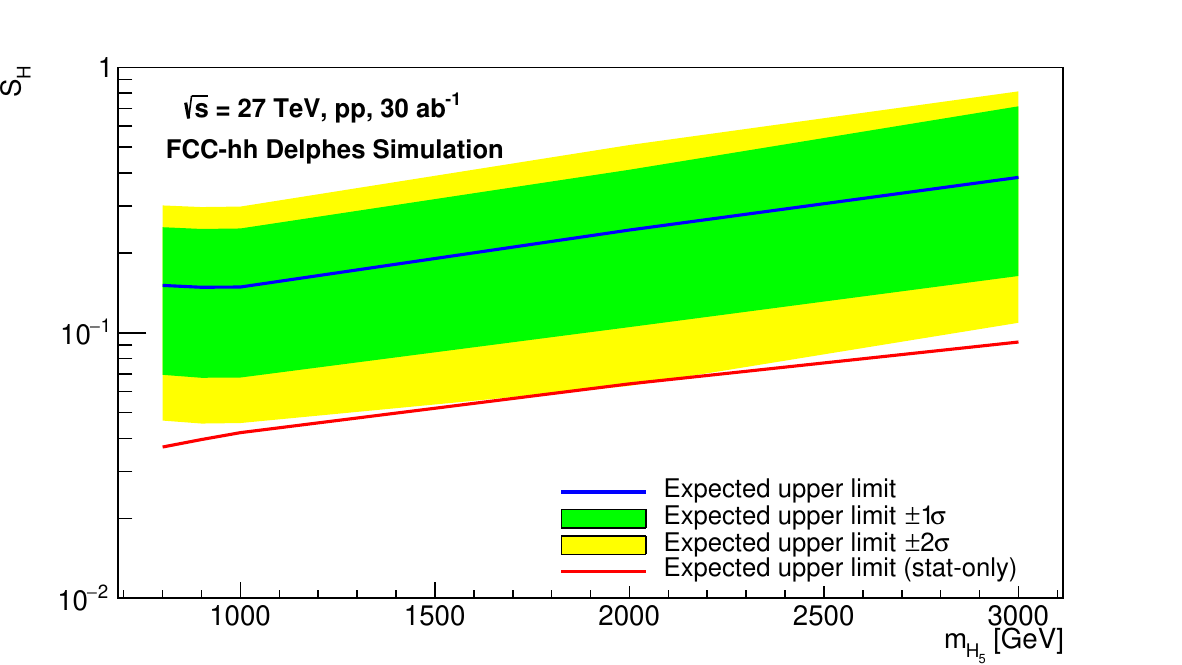}
\includegraphics[width=.49\linewidth]{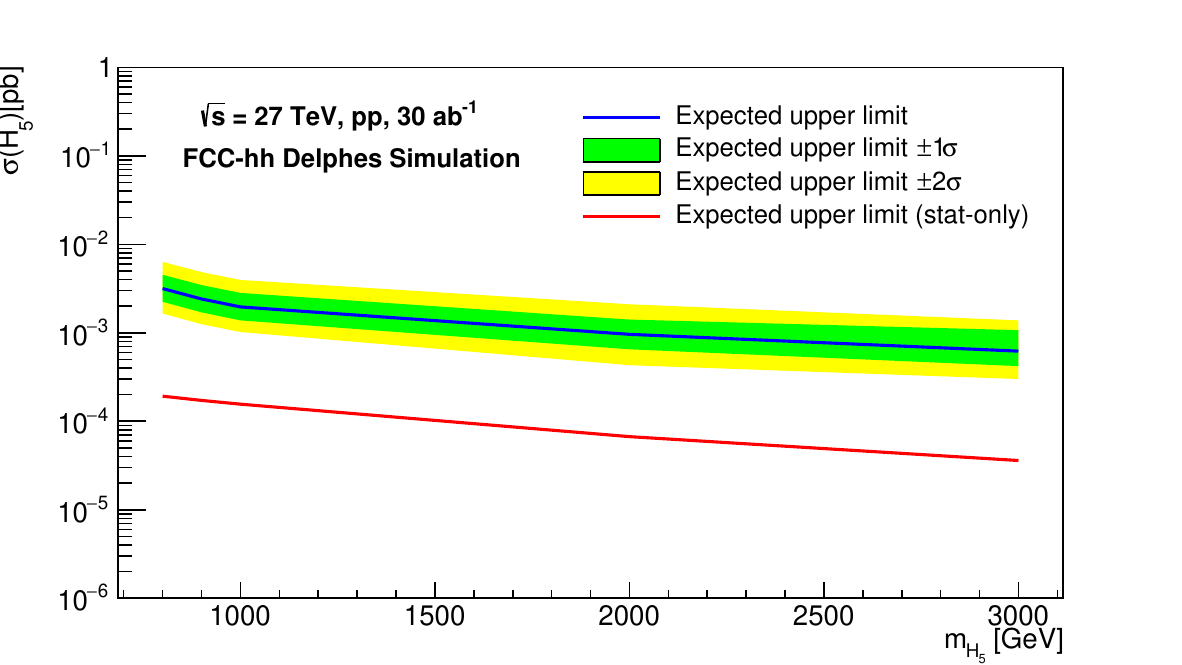}

\includegraphics[width=.49\linewidth]{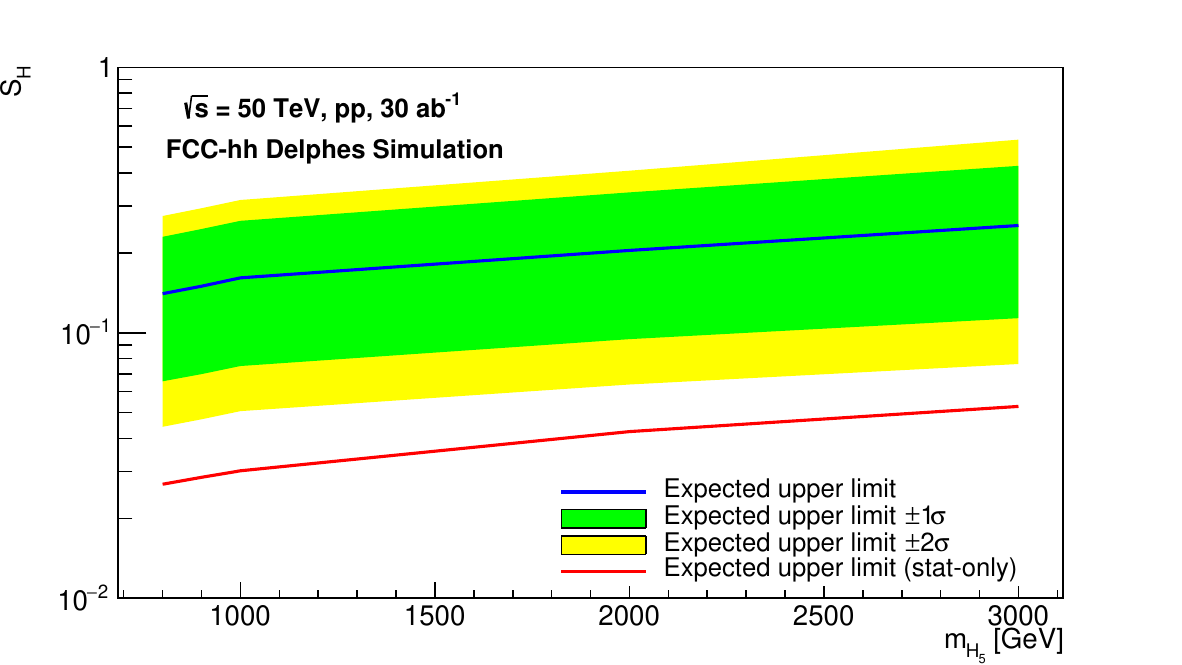}
\includegraphics[width=.49\linewidth]{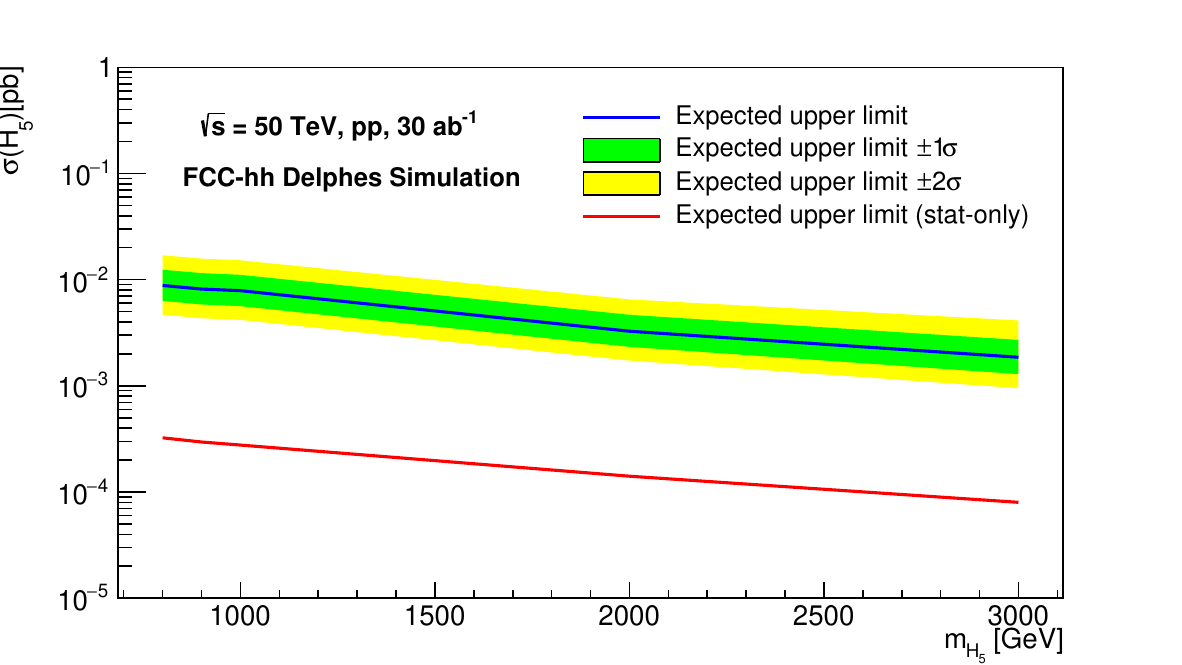}

\includegraphics[width=.49\linewidth]{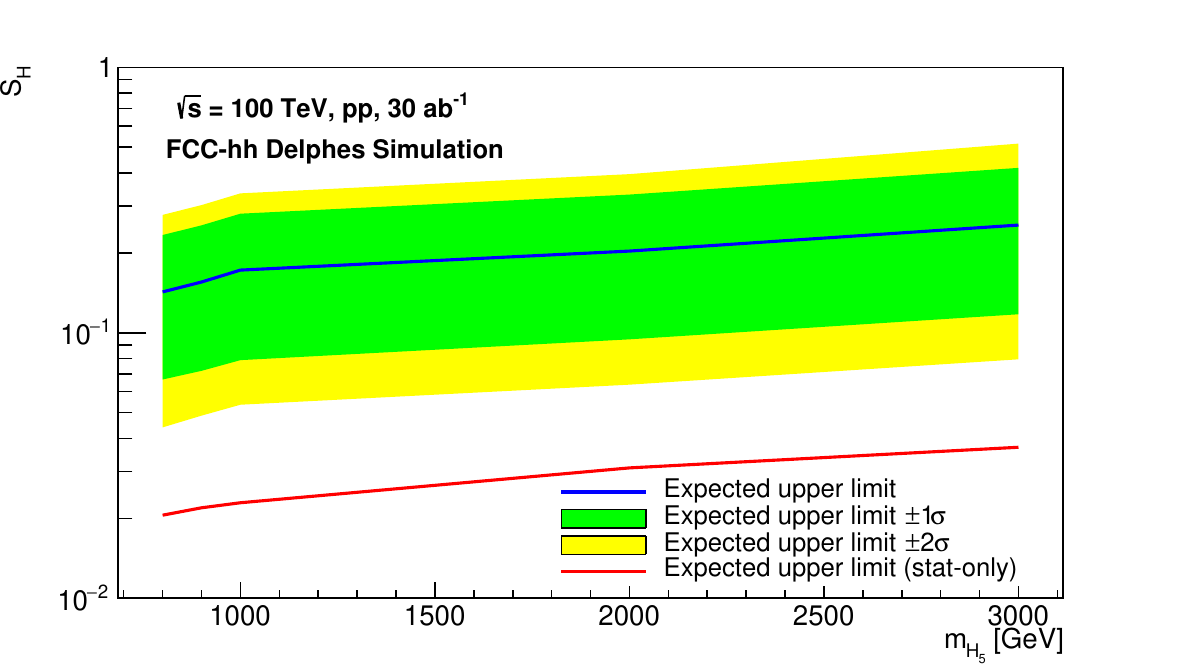}
\includegraphics[width=.49\linewidth]{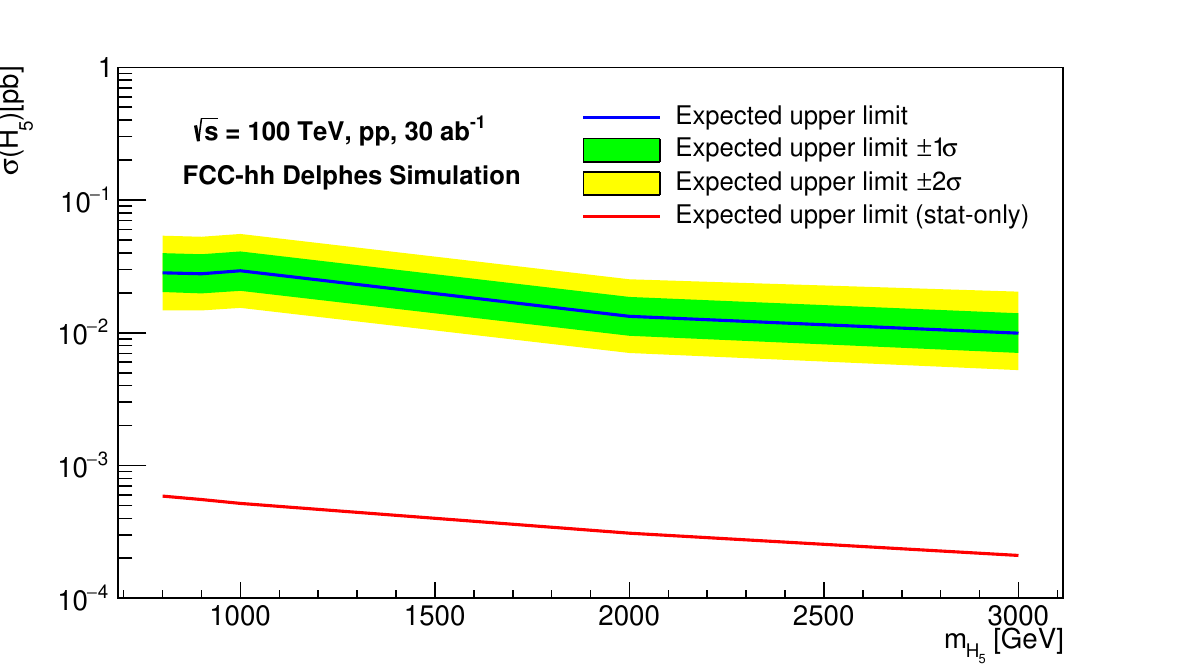}
\caption{Expected exclusion limits at $95\%$ CL on $s_H$ (left) and  $\sigma(H^{\pm\pm}_5)\times \mathcal{B}(H^{\pm\pm}_5\rightarrow W^{\pm}W^{\pm})$ (right) as a function of $m_{H^{\pm\pm}_5}$ at the three center of mass energy scenarios: 27 TeV (top), 50 TeV (center) and 100 TeV (bottom). The blue solid line indicates exclusion limits with both statistical and systematic uncertainties included whereas the red solid line indicates exclusion limits without systematic uncertainties. The green and yellow bands indicate the $68\%$ and $95\%$ confidence intervals around the median limits, respectively.}
\label{Hpp:limits}
\end{figure}

Figure \ref{Hpp:postfit} shows $m_T$ post-fit distributions for the 2000 GeV $H^{\pm\pm}_5$ mass point in the signal region. Expected upper limits on the $s_H$ parameter as a function of $m_{H^{\pm\pm}_5}$ are shown in Figure \ref{Hpp:limits}. Upper limits with and without systematic uncertainties are shown. In general, for the 100 TeV center-of-mass energy, these upper limits range between 0.14-0.25 (with systematic uncertainties) and 0.02-0.04 (without systematic uncertainties). The strong expected upper limits in the latter case is mainly due to the statistical uncertainties arising from the limited number of MC events. In addition, expected upper limits on $\sigma(H^{\pm\pm}_5)\times \mathcal{B}(H^{\pm\pm}_5\rightarrow W^{\pm}W^{\pm})$ are also shown.

 \FloatBarrier
\section{\label{sec:conclusions}Conclusions}

Cross section measurements of the polarized scattering of same-sign \ssWWVBS~production using the leptonic decay mode at a
future high energy $pp$ collider with an integrated luminosity of 30 ab$^{-1}$ should be possible with a relative 
precision of around 22\% at $\sqrt{s} = 27$~TeV,  21\% at $\sqrt{s} = 50$~TeV and 15\% at $\sqrt{s} = 100$~TeV 
for the purely longitudinal contribution and better for the other contributions. An interpretation of the results in the context of the Georgi-Machacek model yields expected exclusion limits on the $s_H$ parameter as a function of $m_{H^{\pm\pm}_5}$. In general, for the 50 TeV and 100 TeV center of mass energies, $s_H$ parameter values greater than 0.14-0.25 are expected to be excluded with the results limited by the number of MC events available for this study.

The effects of NLO EW corrections are not considered for the \ssWWVBS~production. The effects of theoretical systematic uncertainties are taken into account, while detector-specific experimental uncertainties are not included and the systematic uncertainties due to the limited number of MC events are not negligible. The baseline FCC-hh detector parameterization within the Delphes framework is used and a conservative event selection is employed
to mitigate the impact from pile-up and detector-specific background processes which are
not incorporated in this study.

\begin{acknowledgments}
The work of A.A. is supported by the U.S. Department of Energy, Office of Science, Office of High Energy Physics under contract no. DE-SC0013542. The work of C.M. and M.-A.P. is supported by the U.S. Department of Energy, Office of Science, Office of High Energy Physics under contract no. DE-SC0012704. The work of K.P. and L.N. is supported by the U.K. Research and Innovation Science and Technology Facilities Council, under project no.  ST/T004568/1 and 2422336, respectively.
\end{acknowledgments}

\bibliographystyle{ieeetr}
\bibliography{ssww}

\end{document}